\documentclass[aps,preprint,amssymb,12pt,floatfix]{revtex4}
\usepackage[pdftex]{graphicx}
\usepackage{lscape,graphicx}
\usepackage{rotating}
\usepackage{epstopdf}
\usepackage{color}
\usepackage{amsmath,amsthm}
\usepackage{wrapfig}

\begin{document}
\title{Extracting folding landscape characteristics of biomolecules using mechanical forces}
\author{Changbong Hyeon}
\affiliation{Korea Institute for Advanced Study, Seoul 130-722, Korea}
\author{Michael Hinczewski and D. Thirumalai}
\affiliation{Biophysics Program, Institute for Physical Sciences and Technology, University of Maryland, College Park, MD 20742, USA}
\date{\today}

\begin{abstract}
In recent years single molecule force spectroscopy has opened a new avenue to provide profiles of the complex energy landscape of biomolecules. 
In this field, quantitative analyses of the data employing sound theoretical models, have played a major role in interpreting data and anticipating outcomes of experiments.  
Here, we explain how by using temperature as a variable in mechanical unfolding of biomolecules in force spectroscopy, the roughness of the energy landscape can be measured without making any assumptions about the underlying reaction coordinate. 
Estimates of other aspects of the energy landscape such as free energy barriers or the transition state
(TS) locations could depend on the precise model used to analyze the experimental data. 
We illustrate the inherent difficulties in obtaining the transition state location from loading rate or force-dependent unfolding rates. 
Because the transition state moves as the force or the loading rate is varied, which is reminiscent of the Hammond effect, it is in general difficult to invert the experimental data. 
The independence of the TS location on force holds good only for brittle or hard biomolecules whereas the TS location changes considerably if the molecule is soft or plastic. 
Finally, we discuss the goodness of the end-to-end distance (or pulling) coordinate of the molecule as a surrogate reaction coordinate in a situation like force-induced ligand-unbinding from a gated molecular receptor as well as force-quench refolding of an RNA hairpin.
\end{abstract}

\maketitle

\section{Introduction}
The energy landscape, which projects the multidimensional conformational space of biopolymers onto low dimension, has played a critical role in visualizing their folding routes \cite{DillNSB97,OnuchicCOSB04,Thirum05Biochem}.  
To account for the rapid, reversible folding and unfolding of proteins, it is suspected that the folding energy landscape of many evolved proteins is relatively smooth, which allows for an efficient navigation of the landscape. 
To be more precise, the gradient of the energy landscape $\Delta F(\chi)$ towards the native basin of attraction (NBA), corresponding to the driving force, is ``large" enough that during the folding process the biomolecule does not get kinetically trapped in local minima (competing basins of attraction (CBA)) for arbitrarily long times. 
Here, $F(\chi)$ is expressed as a function of a non-unique variable, namely, 
the structure overlap function $\chi$, 
an order parameter that measures how similar a given conformation is to the native state.  
However, perfectly smooth energy landscapes are difficult to realize because of energetic and topological 
frustration \cite{Thirumalai96ACR,Clementi00JMB}. 
In proteins, the hydrophobic residues prefer to be sequestered in the interior 
while polar and charged residues are better accommodated on the surfaces 
where they can interact with water. 
Often these conflicting requirements cannot be simultaneously satisfied and hence proteins can be energetically ``frustrated''. 
It is clear from this description that only evolved or well designed sequences can minimize 
energetic frustration. Even if a particular foldable sequence minimizes energetic conflicts, it is 
nearly impossible to eliminate topological frustration which arises due to 
chain connectivity \cite{Guo95BP,Thirumalai00RNA}. 
Topological frustration refers to the conflict between local and global packing of structures. 
Both sources of frustration, energetic and topological, render the energy landscape rugged on length scales that are larger than those associated with secondary structures ($\approx (1-2)$ nm). 

Thus, the free energy, projected along  a 1D coordinate, is rough on certain length scale and may be globally smooth on a larger scale. 
Under the assumption that the characteristic roughness $\overline{\epsilon}$ has a Gaussian distribution ($P(\overline{\epsilon})\sim e^{-\overline{\epsilon}^2/2\epsilon^2}$), 
the overall transition time from the unfolded basin to NBA over free energy barrier $\Delta F^{\ddagger}$ and $\tau(\beta)=\tau_oe^{\beta\Delta F^{\ddagger}}$ may be written as 
\begin{equation}
\tau_{U\rightarrow F}=\tau(\beta)\int^{\infty}_{k_BT}d\overline{\epsilon}e^{\beta\overline{\epsilon}}P(\overline{\epsilon})\approx \tau(\beta)\int^{\infty}_{k_BT}d\overline{\epsilon}
e^{\beta\overline{\epsilon}}e^{-\overline{\epsilon}^2/2\epsilon^2}\approx
\tau(\beta)e^{\beta^2\epsilon^2/2}
\label{eqn:roughness}
\end{equation}
where $\tau(\beta)$ is the temperature dependent Arrehius-like transition time from unfolded (U) to folded state (F) over a single barrier in the perfectly smooth one-dimensional coordinate, and $\epsilon$ is the average value of ruggedness. 
The last part of the equation becomes valid at low temperatures. 
The additional factor $e^{\beta^2\epsilon^2/2}$ in Eq.\ref{eqn:roughness}, which slows down the folding time, was derived in an elegant paper by 
Zwanzig \cite{ZwanzigPNAS88} and was also obtained in \cite{Bryngelson89JPC,ThirumalaiPRA89} by analyzing the dynamics of Derrida's random energy model \cite{Derrida81PRB}. 
If folding takes place in a rough energy landscape then the dependence of the characteristic time scale on length $l$ may be estimated as 
$\tau=\tau_{SS}\approx (10-100)$ ns when $l\approx (1-2)$ nm where $\tau_{SS}\approx l^2/D$ is the diffusion time with a diffusion constant on the order of $(10^{-7}-10^{-6})$ cm$^2$/sec, and $\tau=\tau_{U\rightarrow F}$ when $l\approx L$ where $L$ is the effective contour length of the biomolecule. 
Given the crude physical picture, the estimate of $\tau_{SS}$ is not inconsistent with the time needed to form $\alpha$-helices or $\beta$-hairpin.

With the possibility of mechanically manipulating biological molecules, one molecule at a time, it is becoming possible to probe the features of their energy landscape (such as roughness and the transition state location) that are not easy to measure using conventional experiments. 
Such experiments, performed using Laser Optical Tweezers (LOTs) \cite{TinocoARBBS04,Ritort06JPHYS} or Atomic Force Microscopy (AFM) \cite{FernandezTIBS99}, have made it possible to mechanically unfold proteins \cite{Marqusee05Science,FernandezNature99,Dietz04PNAS,GaubSCI97,BustamantePNAS00,Oberhauser98Nature,Bustamante94SCI}, RNA \cite{Bustamante01Sci,Bustamante03Science,Woodside06PNAS,Block06Science,TinocoBJ06,Tinoco06PNAS,OnoaCOSB04}, and their complexes \cite{Moy94Science,Fritz98PNAS,EvansNature99,Schwesinger00PNAS,Zhu03Nature,NevoNSB03}, or initiate refolding of proteins \cite{Fernandez04Science} and RNA \cite{TinocoBJ06,Tinoco06PNAS}. 
These remarkable experiments show how the initial conditions affect refolding and also enable us to examine the response of biological molecules over a range of forces and loading rates. In addition, fundamental aspects of statistical mechanics, including non-equilibrium work theorems \cite{JarzynskiPRL97,Crooks99PRE}, can be rigorously tested using single molecule experiments \cite{Bustamante02Science,Trepagnier04PNAS}. Here, we are concerned with using the data and theoretical models to extract key characteristics of the energy landscape of biological systems.

The crude physical picture of folding in a rough energy landscape (Fig.\ref{landscape}) is not meaningful unless the ideas can be validated experimentally, which requires direct measurement of the roughness energy scale $\epsilon$, absolute value of the barrier height, etc.
In conventional experiments, in which folding is triggered by temperature, 
it is difficult to measure $\tau_0$ and $\Delta F^{\ddagger}$ even when $\beta\epsilon\equiv 0$ \cite{Yang03Nature}. 
We proposed, using theoretical methods, that $\beta\epsilon$ can be directly measured using forced-unfolding of biomolecules and biomolecular 
complexes. 
The Hyeon-Thirumalai (HT) theory \cite{Hyeon03PNAS} showed that if unbinding or unfolding lifetime (or rates) are known as a function of the stretching force ($f$) and temperature ($T$) then $\epsilon$ can be inferred without explicit knowledge of $\tau_0$ or $\Delta F^{\ddagger}$. Recently, the loading-rate dependent unbinding times of a protein-protein complex using atomic force microscopy (AFM) at various temperatures have been used to obtain an estimate of $\epsilon$ \cite{Reich05EMBOrep}. Similarly, variations in the forced-unfolding rates as a function of temperature of \emph{Dictyostelium discoideum filamin} (ddFLN4) were used to
estimate $\epsilon$ \cite{RiefJMB05}. 
The variation in unbinding or unfolding rates of proteins as a function of $f$ and $T$ provides an opportunity to obtain quantitative estimates of the energy landscape characteristics.

Single molecule force spectroscopy can also be used to measure force-dependent unfolding rates from which the location of the transition state (TS) in terms of the spatial extension ($R$) can be computed.  
This procedure is a not straightforward because, as shown in a number of studies \cite{Hyeon06BJ,Hyeon05PNAS,Lacks05BJ,West06BJ,Ajdari04BJ}, the location of the transition state changes as $f$ changes unless the curvature of the free energy profile at the TS location is large, i.e, the barrier is sharp. 
The extent to which the TS changes depends on the load.  
By carefully considering the variations of force distributions it is possible 
to obtain reliable estimates of the TS location \cite{RiefJMB05}. 
Here, we review recent developments in single molecule force spectroscopy that have attempted to obtain the energy landscape characteristics of biological molecules \cite{Bell78SCI,Evans97BJ,Hyeon03PNAS,Dudko03PNAS,HummerBJ03,Barsegov05PRL,Barsegov06BJ,Ajdari04BJ}. Using theoretical models that consider dynamics in higher dimensions we also  point out some of the ambiguities in interpreting the experimental data from dynamic force spectroscopy.

\section{Theoretical background}
Single molecule mechanical unfolding experiments differ from conventional 
unfolding experiments in which unfolding (or folding) is triggered by varying temperature or concentration of denaturants or ions. 
In single molecule experiments folding or unfolding can be initiated by precisely manipulating the initial conditions. 
In both forced-unfolding and force-quench refolding, the initial conformation, characterized by the extension of biomolecule, is precisely known. By contrast, the nature of the unfolded states, from which refolding is initiated, is hard to describe in ensemble experiments.  
For RNA and proteins, whose energy landscape is complex \cite{Treiber01COSB,Thirum05Biochem}, details of the folding pathways can be directly monitored by probing the time dependent changes in the end-to-end distance $R(t)$ of individual molecules.  
Analysis of such mechanical folding and unfolding trajectories allows one to explore regions of the energy landscape that are difficult to probe using ensemble experiments.

Force experiments can measure the extension of the molecule as a function of time.  
There are three modes in which stretching experiments are performed. 
Most of the initial experiments were performed by unfolding biomolecules (especially proteins) by pulling on one of the molecule at a constant velocity while keeping the other end fixed \cite{Bustamante01Sci,Bustamante03Science,FernandezNature99,Fisher00NSB}. 
More recently, it has become possible to apply constant force on the molecule of interest using feed-back 
mechanism \cite{Visscher99Nature,Schlierf04PNAS,Fernandez04Science,TinocoBJ06,Fernandez06NaturePhysics}. 
In addition, force-quench experiments have been reported in which the forces are decreased or increased linearly \cite{Fernandez04Science,TinocoBJ06,Tinoco06PNAS}. 
It is hoped that a combination of such experiments can provide a detailed picture of the complex energy landscape of proteins and RNA.  
In all the modes, the  variable conjugate to $f$ is the natural coordinate that describes 
the progress of the reaction of interest (folding, unbinding or catalysis).
If there is a energy barrier confining the molecular motion to a local minimum, 
whose height is greater than $k_BT$, then a sudden increase (decrease) of extension (force) signifies the transition of the molecule over the barrier. 
A rip  in the force-extension curve (FEC) is the signature of such a transition.   Surprisingly, for proteins and RNA it has been found that the portions of the FEC between rips can be quantitatively fit using the semi-flexible or worm-like chain model \cite{MarkoMacro96,Bustamante97Science,Bustamante01Sci,Bustamante03Science,FernandezNature99,Fisher00NSB}. 
From such fits, the global polymeric properties of the biomolecule, 
such as the contour length and the persistence length can be extracted \cite{MarkoMacro96,Bustamante94SCI}.

Single molecule pulling experiments provide distributions of  the unfolding times (or unfolding force) by varying external conditions.  The objective is to construct the underlying energy landscape from such measurements and from mechanical folding or unfolding trajectories.  However, it is difficult to construct all the features of the energy landscape of biomolecules from FEC or mechanical folding trajectories that report only changes at two points.  For example, although the signature of roughness in the energy landscape may be reflected as fluctuations in the dynamical trajectory  it is difficult to estimate its value unless multiple pulling experiments are performed.  We had proposed that the power of single molecules can be more fully realized if temperature ($T$) is also used as an additional variable \cite{Klimov01JPCB,Klimov00PNAS,Hyeon03PNAS,Hyeon05PNAS}. By using $T$ and $f$ it is possible to obtain a phase diagram as a function of $T$ and $f$  that can be used to probe the nature of collapsed molten globules which are invariably populated but are hard to detect in conventional experiments.  We also showed theoretically that the roughness energy scale ($\epsilon$) can be measured \cite{Hyeon03PNAS} if both $f$ and temperature ($T$) are varied in single molecule  experiments.   The effect of $\epsilon$ manifests itself in the  $1/T^2$ dependence of the rates of force-induced unbinding or unfolding kinetics.  In the following subsections we first review the theoretical framework to describe the force-induced unfolding kinetics and show how the 
$1/T^2$ dependence emerges when the roughness is treated as a perturbation in the underlying energy profile.

{\it Bell model: }
Historically,  phenomenological  descriptions of the force-induced fracture of materials and unbinding of adhesive contacts were made by Zhurkov \cite{Zhurkov65IJFM} and Bell \cite{Bell78SCI}, respectively, long before single molecule experiments were performed. In the context of ligand unbinding from a binding pocket, 
Bell \cite{Bell78SCI} conjectured that the kinetics of bond rupture can  be described using a  modified Eyring rate theory \cite{EyringJCP35}, 
\begin{equation}
k=\kappa\frac{k_BT}{h}e^{-(E^{\ddagger}-\gamma f)/k_BT}
\label{eqn:Erying}
\end{equation} 
where $k_B$ is the Boltzmann constant, $T$ is the temperature, 
$h$ is the Planck constant, and $\kappa$ is the transmission coefficient.  
In the Bell description the activation barrier $E^{\ddagger}$ is reduced by a factor $\gamma f$ when the bond or the biomolecular complex is subject to  
external force $f$.  The parameter  $\gamma$ is  a characteristic length of the system 
under load and specifies the distance at which the molecule unfolds or the ligand unbinds. 
The prefactor $\frac{k_BT}{h}$ is the vibrational frequency of a
single bond. The Bell model shows that the unbinding rates increase when tension is applied to the molecule.
Although Bell's key conjecture, i.e., the reduction of th activation barrier due to external force, is physically justified, 
the assumption that $\gamma$ does not depend on the load is in general not valid. In addition, 
because of the multidimensional nature of the energy landscape of biomolecules, 
there are multiple unfolding pathways which require modification of the Bell description of forced-unbinding.  It is an oversimplification to restrict the molecular response to the force merely to a  reduction in the free energy barrier.
Nevertheless, in the experimentally accessible range of loads the Bell model in conjunction
with Kramers' theory of escape from a potential well have been remarkably successful in fitting much of the data on forced-unfolding of biological molecules. 

{\it Mean first passage times:}
In order to go beyond the popular Bell model many attempts have been made to describe the unbinding process as an escape from a free energy surface in the presence of force \cite{Evans97BJ,HummerBJ03,Hyeon03PNAS,Barsegov05PNAS,Dudko06PRL}.  
This is traditionally achieved by a formal procedure that adapts the  Liouville equation describing the time evolution of the  
probability density representing the molecular configuration in phase space.

For the problem at hand, one can project the entire dynamics onto a single reaction coordinate 
provided the relaxation times of other degrees of freedom are shorter than the time scale associated with the presumed order parameter of interest \cite{Zwanzig60JCP,ZwanzigBook}.  In applications to force-spectroscopy, we assume that the variable conjugate to $f$ is a reasonable approximation to the reaction coordinate. The probability density of the molecular configuration, $\rho(x,t|x_0)$, whose 
configuration is represented by order parameter $x$ at time $t$,  obeys the Fokker-Planck equation.

\begin{equation}
\frac{\partial\rho(x,t|x_0)}{\partial t}=\mathcal{L}_{FP}(x)\rho(x,t|x_0)=\frac{\partial}{\partial x}D(x)\left(\frac{\partial}{\partial x}+\frac{1}{k_BT}\frac{dF(x)}{dx}\right)\rho(x,t|x_0), 
\label{eqn:FP}
\end{equation}
where $D(x)$ is the position-dependent diffusion coefficient, 
and  $F(x)$ is an effective one-dimensional free energy, $x_0$ is the position at time $t=0$. 
If the initial distribution is given by $\rho(x_0,t=0|x_0)=\delta(x-x_0)$ 
the formal solution of the above equation reads $\rho(x,t|x_0)=e^{t\mathcal{L}_{FP}}\delta(x-x_0)$. 
If we use an absorbing boundary condition at a suitably defined location, the 
probability that the molecule remains bound (survival probability)  at time $t$ is
\begin{equation}
S(x_0,t)=\int dx\rho(x,t|x_0)=\int dxe^{t\mathcal{L}_{FP}}\delta(x-x_0). 
\label{eq:S}
\end{equation}
In terms of the first passage time distribution, $p_{FP}(x_0,t)=-dS(x_0,t)/dt$, the
mean first passage time can be computed using,
\begin{eqnarray}
\tau(x_0)&=&\int_0^{\infty}dt \left[t p_{FP}(x_0,t)\right]\nonumber\\
&=&-\int^{\infty}_0dt\left[t\frac{dS(x_0,t)}{dt}\right]\nonumber\\
&=&\int^{\infty}_0dt\int dxe^{t\mathcal{L}_{FP}(x)}\delta(x-x_0)\nonumber\\
&=&\int^{\infty}_0dt\int dx\delta(x-x_0)e^{t\mathcal{L}^{\dagger}_{FP}(x)
}
\label{eq:detail}
\end{eqnarray}
where  $\mathcal{L}^{\dagger}_{FP}$ is the adjoint operator. 
In obtaining the above equation we used $S(x_0,t=\infty)=0$ and integrated by parts in going from the second to third line.
By operating $\mathcal{L}_{FP}^{\dagger}(x_0)$ on both sides of Eq.\ref{eq:detail} and exchanging the variable $x$ with $x_0$ we obtain
\begin{equation}
\mathcal{L}^{\dagger}_{FP}(x)\tau(x)=e^{F(x)/k_BT}\frac{\partial}{\partial x}D(x)e^{-F(x)/k_BT}\frac{\partial}{\partial x}\tau(x)=-1. 
\end{equation}
The rate process with reflecting boundary $\partial_x\tau(a)=0$ and absorbing boundary condition $\tau(b)=0$ in the interval $a\leq x\leq b$, leads to 
the well-known expression for mean first passage time, 
\begin{equation}
\tau(x)=\int^b_x dye^{F(y)/k_BT}\frac{1}{D(y)}\int^y_a dz e^{-F(z)/k_BT}.
\label{eqn:mfpt}
\end{equation}

{\it Diffusion in a rough potential:}
In the above analysis the 1D free energy profile $F(x)$  that 
approximately describes the unfolding or unbinding event is arbitrary.  In 
order to explicitly examine the role of the energy landscape ruggedness we 
follow Zwanzig and  decompose $F(x)$ into $F(x)=F_0(x)+F_1(x)$ \cite{ZwanzigPNAS88}. 
where $F_0(x)$ is a smooth potential that determines  the global shape of the energy landscape, and $F_1(x)$ is periodic ruggedness superimposed on $F_0(x)$. 
By taking the spatial average over $F_1(x)$  using 
$\langle e^{\pm\beta F_1(x)}\rangle_l=\frac{1}{l}\int^l_0dx e^{\pm\beta F_1(x)}$, where $l$ is the ruggedness length scale, the 
 associated mean first passage time can be written in terms of the effective diffusion constant $D^*(x)$ as, 
\begin{eqnarray}
D^*(x)&=&\frac{D(x)}{\langle e^{\beta F_1(x)}\rangle_l\langle e^{-\beta F_1(x)}\rangle_l}, \nonumber\\
\tau(x)&\approx&\int^b_x dye^{F_0(y)/k_BT}\frac{1}{D^*(y)}\int^y_a dz e^{-F_0(z)/k_BT}.
\label{eq:MFPT}
\end{eqnarray}
 An inversion of roughness potential, i.e., $F_1\leftrightarrow -F_1$ does not 
alter $D^*(x)$. In the presence of roughness $D^*(x)\leq D(x)$.
When $\beta^2\langle F_1^2(x)\rangle=\beta^2\epsilon^2$ is small the effective diffusion coefficient can be approximated as $D_0^*\approx D_0\exp{\left(-\beta^2\epsilon^2\right)}$ where $D_0$ is the bare diffusion constant. If $P(F_1)$ is a Gaussian then this expression is exact. 
The coefficient associated with $1/T^2$ behavior is due to the energy landscape roughness provided the extension is a good reaction coordinate.
The dependence of $e^{-\epsilon^2/k_B^2T^2}$ in $D^*$ suggests that the diffusion in a rough potential can be substantially slowed even when the scale of roughness is not too large.

{\it Barrier crossing dynamics in a tilted potential:} In writing the one-dimensional Fokker-Planck equation (Eq.\ref{eqn:FP}) we 
assumed that the order parameter $x$ is a slowly changing variable. 
This assumption is valid if the molecular extension, in the presence of $f$, describes accurately the conformational changes  in the biomolecule.

Following the Bell's conjecture we can replace $F(x)$ by $F(x)-f x$ in which $f$ "tilts" the free energy surface.  Thus,  in the presence of mechanical force Eq.\ref{eq:MFPT} becomes 
\begin{equation}
k^{-1}(f)=\tau(f;x)\approx\int^b_x dye^{(F_0(y)-fy)/k_BT}\frac{1}{D^*(y)}\int^y_a dz e^{-(F_0(z)-fz)/k_BT}. 
\label{eqn:MFPT_force}
\end{equation}
As long as the energy barrier is large enough (see Fig.\ref{landscape})
Eq.\ref{eqn:MFPT_force} can be further simplified using the saddle point approximation. 
The Taylor expansions of the free energy potential $F_0(x)-fx$ at the barrier top and the minimum result in Kramers' equation \cite{Kramers40Physica,Hanggi90RMP}, 
\begin{eqnarray}
k^{-1}(f)=\tau(f)&\approx&\frac{2\pi k_BT}{D^*m\omega_b(f)\omega_{ts}(f)}e^{\beta(\Delta F_0^{\ddagger}(f)-f\Delta x^{\ddagger}(f))}\nonumber\\
&=&\left(\frac{2\pi\zeta}{\omega_b(f)\omega_{ts}(f)}\langle e^{\beta F_1(x)}\rangle_l\langle e^{-\beta F_1(x)}\rangle_l\right)e^{\beta(\Delta F_0^{\ddagger}(f)-f\Delta x^{\ddagger}(f))}
\label{eqn:Kramers}
\end{eqnarray}
where $\omega_b$ and  $\omega_{ts}$ are the curvatures of the potential, $|\partial^2_xF_0(x)|$, at $x=x_b$ and $x_{ts}$, respectively, the free energy barrier
$\Delta F_0^{\ddagger}=F_0(x_{ts})-F_0(x_b)$, $m$ is the effective mass of the biomolecule,
$\zeta$ is the friction coefficient, and $\Delta x^{\ddagger}\equiv x_{ts}-x_b$.

In the presence of $f$, the positions of the transition state $x_{ts}$ and bound state $x_b$ change because unbinding
kinetics should be  determined using $F_0(x)-fx$ and not $F_0(x)$ alone. Because 
$x_{ts}$ and $x_b$ satisfy the \emph{force dependent condition} $F_0'(x)-f=0$, 
it follows that all the parameters, $\Delta x^{\ddagger}(f)$, $\omega_{ts}(f)$, and $\omega_b(f)$,  are intrinsically $f$-dependent. 
Depending on the shape of the free energy potential $F_0(x)$, the degree of 
force-dependence of $\Delta x^{\ddagger}$, $\omega_{ts}$, and $\omega_b$ can vary greatly. 
Previous theoretical studies \cite{HummerBJ03,Dudko03PNAS,Hyeon06BJ,RitortPRL06} have examined some of the consequences of the moving transition state.  
In addition, simulational studies \cite{Hyeon05PNAS,Lacks05BJ,Hyeon06BJ} in which  the free energy profiles  were explicitly computed from thermodynamic considerations alone clearly showed the 
change of $\Delta x^{\ddagger}$ when $f$ is varied.  These authors also provided a structural basis for transition state movements in the case of unbinding of simple RNA hairpins. 
The nontrivial coupling of force and free energy profile makes it difficult to 
unambiguously extract free energy profiles from experimental data. 
In order to circumvent some of the problems, Schlierf and Rief have used Eq.\ref{eqn:MFPT_force}  to analyze the load-dependent experimental data on unfolding of ddFLN4 and extracted an effective
one dimensional free energy surface $F(x)$ without making additional assumptions.  The results showed that the effective free energy profile is highly anharmonic near the transition state region \cite{Schlierf06BJ}.

{\it Forced-unfolding dynamics at constant loading rate: } Many single molecule experiments are conducted by ramping the force over time \cite{Bustamante02Science,Bustamante03Science,FernandezNature99,FernandezTIBS99}. In this mode the load on the molecule or the complex increases with time. When the force increases beyond a threshold value, unbinding or bond-rupture occurs. Because of thermal fluctuations the unbinding events are stochastic and as a consequence one has to contend with the distribution of unbinding forces.  The time-dependent nature of the force makes the barrier crossing rate also dependent on $t$. 
For a single barrier crossing event with a time-dependent rate $k(t)$, 
the probability of the barrier crossing event being observed at time $t$ is 
$P(t)=k(t)S(t)$, where the survival probability $S(t)$ that the molecule remains 
folded is given as $S(t)=\exp{\left(-\int ^t_0d\tau k(\tau)\right)}$.

When the molecule or complex is pulled at a  constant loading rate ($r_f$) 
the distribution ($P(f)$) of unfolding forces is asymmetric.  The most probable $r_f$-dependent unfolding force ($f^*$) is often used to determine  the TS location of the underlying energy landscape with the tacit assumption that the TS is stationary. When $r_f=df/dt$ is constant, the probability of observing an  unfolding event at force $f$ is written as, 
\begin{equation}
P(f)=\frac{1}{r_f}k(f)\exp{\left[-\int^f_0df'\frac{1}{r_f}k(f')\right]}. 
\label{eqn:force_distribution}
\end{equation} 
The most probable unfolding force is obtained from
$dP(f)/df|_{f=f^*}=0$, which leads to
\begin{eqnarray} 
f^*&=&\frac{k_BT}{\Delta x^{\ddagger}(f^*)}\lbrace\log{\left(\frac{r_f\Delta x^{\ddagger}(f^*)}{\nu_D(f^*)e^{-\beta\Delta F^{\ddagger}_0(f^*)}k_BT}\right)}\nonumber\\ 
&+&\log{\left(1+f^*\frac{(\Delta x^{\ddagger})'(f^*)}{\Delta x^{\ddagger}(f^*)}-\frac{\left(\Delta F^{\ddagger}_0\right)'(f^*)}{\Delta x^{\ddagger}(f^*)}
+\frac{\nu'_D(f^*)}{\nu_D(f^*)}\frac{k_BT}{\Delta x^{\ddagger}(f^*)}\right)}\nonumber\\ 
&+&\log{\langle e^{\beta F_1}\rangle_l\langle e^{-\beta F_1}\rangle_l}\rbrace,
\label{eqn:most_force}
\end{eqnarray}
where $\Delta F^{\ddagger}_0\equiv F_0(x_{ts}(f))-F_0(x_0(f))$, $\prime$ 
denotes differentiation with respect to the argument, $\Delta x(f)\equiv x_{ts}(f)-x_0(f)$ is the distance between the transition state and the native state, and 
$\nu_D(f)\equiv \omega_o(f)\omega_{ts}(f)/2\pi\gamma$. 
Note that $\Delta F^{\ddagger}_0$, $\nu_D$, and $\Delta x^{\ddagger}$ depend on the value of $f$ \cite{Hyeon03PNAS,Lacks05BJ,Hyeon05PNAS}.  
Because $f^*$ changes with $r_f$, $\Delta x^{\ddagger}$ obtained from the data analysis should correspond to a value at a certain $f^*$, not a value that is extrapolated to $f^*=0$.  Indeed, the pronounced curvature in the plot of $f^*$ as a function of $\log{r_f}$  makes it difficult to obtain the characteristics of the underlying energy landscape using data from dynamic force-spectroscopy without a reliable theory or a model.
If $\Delta F^{\ddagger}_0$, $\nu_D$, and $\Delta x^{\ddagger}$ are relatively insensitive to variations in  force, the second term on the right-hand side of Eq.\ref{eqn:most_force} would vanish, leading to $f^*\propto (k_BT/\Delta x^{\ddagger})\log{r_f}$ \cite{Evans97BJ}. 
If the loading rate, however, spans a wide range so that the force-dependence of $\Delta F^{\ddagger}_0$, $\nu_D$, and $\Delta x^{\ddagger}$ are manifested, then the resulting $f^*$ can substantially deviate from the linear dependence on $\log{r_f}$.  Indeed, it has been shown that for a molecule or a complex known to have a single free energy barrier, the most probable rupture force $f^*$ obeys $(1-f^*/f_c)\sim (\log {r_f})^{\nu}$ with $f_c$ being a critical force in the absence of force. 
The effective exponent $\nu$ should lie in the range $0.5 \le \nu \le 1$ \cite{Hyeon2012JCP}.  
And a fit with $\nu<0.5$ implies that the unbinding dynamics cannot be explained with a one-dimensional barrier picture of crossing \cite{Hyeon2012JCP}. 
The precise value of $\nu$ depends on the nature of the underlying potential and is best treated as an adjustable parameter.

\section{Measurement of energy landscape roughness}
In the presence of roughness we expect that the unfolding kinetics deviates substantially from an Arrhenius behavior.
By either assuming a Gaussian distribution of the roughness contribution, $P(F_1)\propto e^{-F_1^2/2\epsilon^2}$, or simply assuming $\beta F_1\ll 1$ and $\langle F_1\rangle=0$, $\langle F_1^2\rangle=\epsilon^2$, one can further simplify 
Eq.\ref{eqn:Kramers} to 
\begin{equation}
\log{k(f)/k_0}=-(\Delta F_0^{\ddagger}-f\cdot\Delta x^{\ddagger})/k_BT-\epsilon^2/k_B^2T^2.
\label{eqn:k_mod}
\end{equation}
This relationship suggests that the roughness scale $\epsilon$ can be extracted if $\log{k(f)}$ is measured over a range of temperatures. Variations in temperature also result in changes in the 
viscosity, $\eta$, and 
because $k_0^{-1}\propto\eta$, corrections arising from the temperature-dependence of $\eta$ have to be taken into account in
interpreting the experiments.
It is known that $\eta$ for water varies as $\exp(A/T)$ over the experimentally relevant temperature range ($5^oC<T<50^oC$) \cite{CRC}. 
Thus, we expect $\log{k(f,T)}=a+b/T-\epsilon^2/T^2$. 
The coefficient of the $1/T^2$ term can be quantified by performing \emph{force-clamp experiments} at several values of constant temperatures. 
In addition, the robustness of the HT theory can be confirmed by showing that $\epsilon^2$ is a constant even if the  coefficients $a$ and $b$ change under 
different force conditions \cite{Hyeon03PNAS}. The signature of the roughness of the underlying energy landscape is uniquely reflected in the
non-Arrhenius temperature dependence of the unbinding rates. Although it is most straightforward to extract $\epsilon$ using Eq.\ref{eqn:k_mod},
no roughness measurement, to the best of our knowledge, has been performed using force clamp experiments.

To extract the roughness scale, $\epsilon$, using dynamic force spectroscopy (DFS) in which the force increases gradually in time, an alternative but similar 
strategy as in force clamp experiments can be adopted.
A series of dynamic  force spectroscopy experiments should be performed as 
a function of $T$ and $r_f$ so that reliable unfolding force distributions are 
obtained. 
Since a straightforward application of Eq.\ref{eqn:most_force} is difficult due to the force-dependence of the variables in Eq.\ref{eqn:most_force}, one should simplify the expression by assuming that the parameters $\Delta x^{\ddagger}(f)$, $\Delta F_0^{\ddagger}(f)$, and $\nu_D(f)$, depend only weakly on $f$. 
If this is the case then  the second term of Eq.\ref{eqn:most_force} can be neglected and Eq.\ref{eqn:most_force} becomes 
\begin{equation}
f^*\approx\frac{k_BT}{\Delta x^{\ddagger}}\log{r_f}+\frac{k_BT}{\Delta x^{\ddagger}}\log{\frac{\Delta x^{\ddagger}}{\nu_De^{-\Delta F_0^{\ddagger}/k_BT} k_BT}}+\frac{\epsilon^2}{\Delta x^{\ddagger} k_BT}.
\label{eqn:most_force_approx}
\end{equation}
One way of obtaining the roughness scale from experimental data is as follows \cite{Reich05EMBOrep}.  From the $f^*$ vs $\log{r_f}$ curves at two different temperatures, $T_1$ and $T_2$, 
one can obtain $r_f(T_1)$ and $r_f(T_2)$ for which the  $f^*$ values are identical. 
By equating the right-hand side of the expression in Eq.\ref{eqn:most_force} at $T_1$ and $T_2$ 
the scale $\epsilon$ can be estimated \cite{Hyeon03PNAS,Reich05EMBOrep} as 
\begin{small}
\begin{eqnarray}
\epsilon^2 &\approx&\frac{\Delta x^{\ddagger}(T_1)k_BT_1\times\Delta x^{\ddagger}(T_2)k_BT_2}{\Delta x^{\ddagger}(T_1)k_BT_1-\Delta x^{\ddagger}(T_2)k_BT_2}\nonumber\\
&\times& \left[\Delta F^{\ddagger}_0\left(\frac{1}{\Delta x^{\ddagger}(T_1)}-\frac{1}{\Delta x^{\ddagger}(T_2)}\right)+\frac{k_BT_1}{\Delta x^{\ddagger}(T_1)}\log{\frac{r_f(T_1)\Delta x^{\ddagger}(T_1)}{\nu_D(T_1)k_BT_1}}-\frac{k_BT_2}{\Delta x^{\ddagger}(T_2)}\log{\frac{r_f(T_2)\Delta x^{\ddagger}(T_2)}{\nu_D(T_2)k_BT_2}}\right]. 
\label{eqn:epsilon}
\end{eqnarray}
\end{small}

Nevo et. al. \cite{Reich05EMBOrep} used DFS to measure $\epsilon$ 
for a biomolecular protein complex consisting of nuclear import receptor importin-$\beta$ (imp-$\beta$) and the Ras-like GTPase Ran that is loaded with non-hydrolyzable GTP analogue (Fig.\ref{Nevo_fig}-C).  
The Ran-imp-$\beta$ complex was immobilized on a surface and the unbinding forces were measured using AFM at three values of $r_f$ that varied by nearly three orders of magnitude.  
At high $r_f$ the values of $f^*$ increases as $T$ increases. 
At lower loading rates ($r_f\lesssim 2\times 10^3$ pN/s), however, $f^*$ decreases as $T$ increases (see Fig.\ref{Nevo_fig}-B).
The data over distinct temperatures were  used to extract, for the first time, an estimate of $\epsilon$.  The values of $f^*$ at three temperatures (7, 20, 32$^oC$) and Eq.\ref{eqn:epsilon} were used to obtain  $\epsilon\approx 5-6k_BT$.  
Nevo et. al. explicitly showed that the value of $\epsilon$ was nearly the same  from the nine pairs of data extracted from the  $f^*$ vs $\log{r_f}$ curves. 
Interestingly, the estimated value of $\epsilon$ is about $0.2 \Delta F^{\ddagger}_0$ where $\Delta F^{\ddagger}_0$ is the major barrier for unbinding of the complex. 
This shows that for this complex the free energy in terms of a one-dimensional coordinate resembles the profile shown in Fig.\ref{landscape}.  
It is worth remarking that the location of the transition state decreases from 0.44 nm at 7 $^o$C to 0.21 nm at 32 $^o$C.  
The extracted TS movement using the roughness model is consistent with Hammond behavior (see below).

\section{Extracting TS location ($\Delta x^{\ddagger}$) and unfolding rate ($\kappa$) from Dynamic Force Spectroscopy}
The theory of DFS, $f^*\approx\frac{k_BT}{\Delta x^{\ddagger}}\log{r_f}+\frac{k_BT}{\Delta x^{\ddagger}}\log{\frac{\Delta x^{\ddagger}}{\kappa k_BT}}$,  is used to identify the forces that destabilize the bound state of the complex or the folded state of a specific biomolecule.   
A  linear regression  provides the characteristic extension  $\Delta x^{\ddagger}$ at which the molecule or complex ruptures (more precisely $\Delta x^{\ddagger}$ is the thermally averaged distance 
between the bound and the transition state along the direction of the applied force). 
It is tempting to obtain the zero force unfolding rate  $\kappa$ from the intercept with the abscissa. 
Substantial errors can, however, arise in the extrapolated values of $\Delta x^{\ddagger}$ and $\kappa$ to the zero force if $f^*$ vs $\log{r_f}$ is not linear, as is often the case when $r_f$ is varied over several orders of magnitude.  
Nonlinearity of the $[f^*,\log{r_f}]$ curve arises for two reasons. One is due to the complicated molecular response to the external load that results in dramatic variations in $\Delta x^{\ddagger}$.  Like other soft matter,  the extent of the response (or the elasticity) depends on $r_f$ \cite{Hyeon06BJ,Lacks05BJ,West06BJ}.
The other is due to multiple energy barriers that are encountered in the unfolding or unbinding process \cite{EvansNature99}.

If the TS ensemble is broadly distributed along the reaction coordinate then the molecule can adopt diverse structures along the energy barrier depending on the magnitude of the external load.
Therefore, mechanical force should grasp the signature of the spectrum of the TS conformations for such a molecule. Mechanical unzipping dynamics of RNA hairpins whose stability is determined in terms of the number of intact base pairs is a good example. 
The conformation of RNA hairpins at the barrier top 
can gradually vary from an almost fully intact 
structure at small forces to an extended structure at large forces.  
Under these conditions the width of the TSE is large. 
The signature of diverse TS conformations manifests itself as a substantial 
curvature over the broad variations of forces or loading rates. 
Meanwhile, if the unfolding is a highly cooperative all-or-none process characterized by a narrow distribution of the TS, the nature of the TS may not change significantly.

The linear theory of DFS is not reliable if the TSE is plastic because it involves  drastic approximations of the Eq.\ref{eqn:most_force}. 
From this perspective it is more prudent to fit the  the experimental unbinding force distributions directly using  analytical expressions derived from suitable models. 
If such a procedure can be reliably implemented then  
the extracted parameters are likely to be more accurate. 
Solving such an inverse problem does require assuming a reduced dimensional representation of the underlying energy landscape which cannot be \textit{a priori} justified. 

\section{Mechanical response of hard (brittle)  versus soft (plastic) biomolecules}

With few exceptions \cite{Barsegov05PNAS}, lifetimes of a complex decrease upon application of force. The compliance of the molecule is determined by the location of the TS, and hence it is important to understand the characteristics of the molecule that determine the TS.  As we pointed out, many relevant paramters have strong dependence on $f$, $r_f$, or $T$. 
Thus, it is difficult to extract the energy landscape parameters without a suitable model.  In this section we illustrate  two extreme cases of mechanical response \cite{West06BJ,Hyeon06BJ,RitortPRL06,West06BJ} of a biomolecule using one-dimensional energy profiles.  In one example the location of the TS does not move with force whereas in the other there is a dramatic movement of the TS. In the presence of force $f$, a given  free energy profile $F_0(x)$ changes to $F(x)$ =  $F_0(x)-fx$.  The location of the TS at non-zero values of $f$ depends on the shape of barrier in the vicinity of the TS.  Near the barrier ($x \approx x_{ts}$) we can approximate $F_0(x)$ as 
\begin{equation}
F_0(x) \approx F_0(x_{ts})-\frac{1}{2} F_0^{\prime\prime}(x_{ts}) (x - x_{ts})^2+\cdots. 
\label{eqn:expansion}
\end{equation}
In the presence of force the TS location becomes $x_{ts}(f)=x_{ts}-\frac{f}{F_0''(x_{ts})}$.  If the transition barrier in $F_0(x)$ is sharp ($x_{ts}F_0''(x_{ts})\gg f$) then we expect very little force-induced movement in the TS.  We refer to molecules that satisfy this criterion as hard or brittle.  In the opposite limit the molecule is expected to be soft or plastic so that there can be dramatic movements in the TS.  We illustrate these two cases by numerically computing $r_f$-dependent  $P(f)$  using Eq.\ref{eqn:MFPT_force} and Eq.\ref{eqn:most_force} for two model free energy profiles. 

{\it Hard response:} A nearly stationary TS position (independent of $f$) 
is realized if the energy barrier is sharp (Eq.\ref{eqn:expansion}). 
We model $F_0(x)$ using 
\begin{equation}
\begin{array}{ll}
     F_0(x)=-V_0|(x+1)^2-\xi^2| & \mbox{with $x\geq 0$}
\end{array}
\label{eqn:hard}
\end{equation}
where $V_0=28$ pN/nm and $\xi=4$ nm. 
The energy barrier forms at $x=1$ nm and this position does not change much even in the presence of force as illustrated in Fig.\ref{hard_soft.fig}A (top panel).  In dynamic force spectroscopy the free energy profiles drawn at constant force may be viewed as  snapshots at different times. 
The shape of the unbinding force distribution  depends on $r_f$. 
We calculated $P(f)$  numerically using Eq.\ref{eqn:MFPT_force} and Eq.\ref{eqn:most_force} (see the middle panel of Fig.\ref{hard_soft.fig}A). 
Interestingly, a plot of the the most probable force $f^*$ obtained from $P(f)$ does not exhibit any curvature when $r_f$ is varied over
six orders of magnitude (the bottom panel of Fig.\ref{hard_soft.fig}A). 
Over the range of $r_f$ the $[f^*, \log{r_f}]$ plot is almost linear. 
The slight deviation from linearity is due to the force-dependent curvature near the bound state ($\omega_b(f)$). 
From the slope we find that  $\Delta x^{\ddagger}\approx 1$ nm which is expected from Eq.\ref{eqn:hard}. 
In addition,  we obtained from the intercept in Fig.\ref{hard_soft.fig}-C that $\kappa=1.58$ $s^{-1}$.  
The value of $\kappa[\equiv k(f=0)]$ directly computed using 
Eq.\ref{eqn:MFPT_force} is $\kappa=1.49$ $s^{-1}$. 
The two values agree quite well. 
Thus, for brittle response the Bell model is expected to be accurate.  

{\it Soft response:} If the  position of the TS \emph{sensitively} moves with force the biomolecule or the complex is soft or plastic. 
To illustrate the behavior of soft molecules we model the free energy potential in the absence of force using 
\begin{equation}
\begin{array}{ll}
     F_0(x)=-V_0\exp{(-\xi x)} & \mbox{with $x\geq 0$}
\end{array}
\label{eqn:soft}
\end{equation}
where $V_0=82.8$ pN$\cdot$nm and $\xi=4$ (nm)$^{-1}$. 
The numerically computed $P(f)$ and $[f^*, \log{r_f}]$ plots are shown in the middle and bottom panels of Fig.\ref{hard_soft.fig}B, respectively.  
The slope of the $[f^*, \log{r_f}]$ plot is no longer constant but increases continuously as $r_f$ increases. 
The extrapolated value of $\kappa$ to zero $f$ varies greatly depending on the range of $r_f$ used.  Even in the experimentally accessible range of $r_f$ there is curvature in the $[f^*,\log{r_f}]$ plot.  Thus, unlike the parameters ($\Delta x^{\ddagger}$, $\kappa$) in the example of a brittle potential, all the extracted parameters from the force profile  are strongly dependent on the loading rate. 
As a result, in soft molecules the extrapolation to zero force (or minimum loading rate) is not as meaningful as in hard molecules. 
Note how the extracted $\Delta x^{\ddagger}$ (see the inset in the bottom panel of Fig.\ref{hard_soft.fig}B) changes as a function of $r_f$. 
For soft (plastic) molecules, the extracted parameters using the tangent at a certain $r_f^o$ are not the characteristics of the free energy profile in the absence of the load, but reflect the  features for the modified free energy profile tilted by $(f^*)^o$ at $r_f^o$.

In practice, biomolecular systems lie between the two extreme cases (brittle and plastic).  
In many cases the $[f^*, \log{r_f}]$ appears to be linear over a narrow range of $r_f$. 
The linearity in narrow range of $\log{r_f}$, however, does not guarantee the linearity under broad variations of loading rates.  
In order to obtain energy landscape  parameters it is important to perform experiments 
at $r_f$ as low as possible.  
The brittle nature of proteins (lack of change in $\Delta x^{\ddagger}$) inferred from AFM experiments may be the result of a relatively large $r_f$ ($\approx 1,000$ pN/s).  
On the other hand, only by varying $r_f$ over a wide range can the molecular elasticity of proteins and RNA 
be completely described.  
Indeed, we showed that even in simple RNA hairpins the transition 
from plastic to brittle behavior can be achieved by varying $r_f$ \cite{Hyeon06BJ}. 
The load-dependent response may even have  functional significance.

\section{Hammond/anti-Hammond behavior under force and temperature variations}
The qualitative nature of the TS movement with increasing perturbations can often be anticipated using the Hammond postulate which has been successful in not only analyzing a large class of chemical reactions but also in rationalizing the observed behavior in protein and RNA foldings.  
The Hammond postulate states that the nature of TS resembles the least stable species along the reaction pathway. In the context of forced-unfolding it  implies that the TS location should move closer to the native state as $f$ increases.  
In other words $\Delta x^{\ddagger}$ should decrease as $f$ is increased.  Originally the Hammond postulate was introduced to explain chemical reactions involving  small organic molecules \cite{HammondJACS53,LefflerSCI53}.  
Its validity in biomolecular folding is not obvious because  there are  multiple folding or unfolding pathways.  
As a result there is a large entropic component to the folding reaction.  Surprisingly, many folding processes are apparently in accord with the Hammond postulate \cite{Fersht95Biochemistry,Dalby98Biochemistry,Kiefhaber00PNAS}. If the extension is an appropriate reaction coordinate for forced unfolding then deviations from Hammond postulate should be an exception rather than the rule.  Indeed, anti-hammond behavior (movement of the TS closer to more stable unfolded state as $T$ increases) was suggested by Schlierf and Rief \cite{RiefJMB05} based on a model used to analyze the AFM data.  
The simple free energy profiles used in the previous section (Eq.\ref{eqn:hard} and Eq.\ref{eqn:soft}) can be used to verify the Hammond postulate when the external perturbation is either force or temperature. 
First, for the case of hard response the TS is barely affected by force, thus the Hammond or anti-Hammond behavior is not a relevant issue when unbinding is induced by $f$.  
On the other hand, for the case of soft molecules $\Delta x^{\ddagger}$ always decreases with a larger force. 
The positive curvature in $[f^*,\log{r_f}]$ plot is the signature of the classical Hammond-behavior with respect to $f$. 

As long as a one dimensional free energy profile suffices in describing forced-unfolding of proteins and RNA the TS location must satisfy the Hammond postulate.  In general, for a fixed force or $r_f$, $\Delta x^{\ddagger}$ can vary with $T$. The changes in $\Delta x^{\ddagger}$ with temperature can be modeled using $T$-dependent parameters in the potential. To evaluate the consequence of $T$-variations we set  
\begin{equation}
\xi=\xi_0+\alpha(T-300K)
\end{equation} 
for both free energies in Eq.\ref{eqn:hard} and Eq.\ref{eqn:soft}. 
Depending on the value of $\alpha$  the position of the TS can move towards or away from the native state. We set $\alpha=\pm 0.1$ for both the hard and soft cases. The numerically computed   $[f^*, \log{r_f}]$ plots are shown in Fig.\ref{hard_soft.fig}. 
One interesting point is found in the soft molecule that exhibits Hammond behavior.  For wide range of $r_f$, $\Delta x^{\ddagger}$ decreases as $T$ increases. However, the most probable unbinding force $f^*$ at low temperatures can be larger or smaller than $f^*$ at high temperatures depending on the loading rate (see upper-right corner of Fig.\ref{hard_soft.fig}).  A very similar behavior has been observed in the forced-unbinding of Ran-imp-$\beta$ complex \cite{Reich05EMBOrep} (see also Fig.\ref{Nevo_fig}-B).
Although the model free energy profiles can produce a wide range of behavior depending on $T$, $f$, and $r_f$ the challenge is to provide a structural basis for the measurements on biomolecules.

\section{Molecular tensegrity and the transition state.} 

For  $R_{TS}$ (see Fig.\ref{Tensegrity.fig}A for the definitions of $R_{TS}$, $R_U$, $R_F$ on a one-dimensional free energy profile $F(R)$), associated  with  the barrier top of $F(R)$ at $f=f_m$ where $f_m$ is the transition mid-force, to be considered the ``true" transition state, it is necessary to ensure that it is consistent  with other conventional definitions of the transition state ensemble. A plausible definition of the TS is that  the forward (to the unfolded state, $P_{unfold}$) and backward  (to the folded state, $P_{fold}$) fluxes  starting from the transition state on the reaction coordinate should be identical \cite{Klosek91BBPC}. For an RNA hairpin it means that if an ensemble of structures were created starting at $R_{TS}$ then the dynamics in the full multidimensional space would result in these structures reaching the folded and unfolded states with equal probability. 
The number of events reaching $R_F$ and $R_U$ starting from  $R_{TS}$ can be directly  counted if folding trajectories with high temporal resolution exhibiting multiple folding and unfolding transitions at $f=f_m$ can be generated (Fig.\ref{Tensegrity.fig}A). Our coarse-grained simulations of the P5GA RNA hairpin, which were the first to assess the goodness of $R_{TS}$ as a descriptor of the TS, showed that 
starting from $R_{TS}$ the hairpin crosses the TS region multiple times before reaching $R=R_F$ or $R_U$, suggesting that the TS region is broad and heterogeneous.
The transition dynamics of biopolymers occurs on a bumpy folding landscape with fine structure even in the TS region, which implies  there is an internal coordinate determining the fate of trajectory projected onto the $R$-coordinate. 
In accord with this inference we showed that for the P5GA hairpin the TS structural ensemble is heterogeneous (Fig. \ref{Tensegrity.fig}B). More pertinently,  the forward and backward fluxes starting from the structure in TS ensemble (see the dynamics of trajectories starting from $R_{TS}$ in Fig.\ref{Tensegrity.fig}B) do not satisfy the equal flux condition, $P_{fold}=P_{unfold}=0.5$. Thus, from a strict perspective $R$ for a simple hairpin may not be a good reaction coordinate even under tension, implying that $R$ is unlikely to be an appropriate reaction coordinate at $f \approx 0$. 
 
Based on simulations Morrison \emph{et al.} \cite{Morrison11PRL} proposed a fairly general theoretical  criterion to determine if $R$ could be a suitable reaction coordinate. 
The theory uses the concept of tensegrity (tensional integrity), which was introduced by Fuller and developed in the context of biology to describe the stability of networks. The notion of  tensegrity has been used to account for cellular structures \cite{IngberJCS03} and more recently for the stability of globular proteins \cite{Edwards12PLoSCompBiol}, the latter of which made an interesting estimate that the magnitude of inter-residue precompression and pretension, associated with structural integrity, can be as large as a few 100 pN.         
Using $F(R)$ the experimentally measurable molecular tensegrity parameter is defined $s\equiv f_c/f_m=\Delta F^{\ddagger}(f_m)/f_m\Delta R^{\ddagger}(f_m)$, where $\Delta F^{\ddagger}=F(R_{TS})-F(R_F)$ and $\Delta R^{\ddagger}=R_{TS}-R_F$. 
The molecular tensegrity parameter $s$ represents a balance between the compression force ($f_m$) and the tensile force ($f_c$), a building principle in tensegrity systems \cite{Fuller61}.  For hairpins such stabilizing interactions involve favorable base pair formation.
 In terms of $s$ and the parameters characterizing the one dimensional landscape ($f_m$ and $k_u$ in Fig. \ref{Tensegrity.fig}A) an analytic expression for $P_{unfold}$ has been obtained \cite{Morrison11PRL}.
\begin{equation}\label{1}
\qquad P_{unfold}(s) = \begin{cases} 
\frac{1}{2}\frac{1}{1+s} & s\gg 1\\ \frac{1}{2}\left[\frac{\Phi}{1+\Phi}+\frac{32f_m^2(\Phi-1)}{\pi k_u(\Phi+1)^3}s\right]^{-1} & s\ll 1 
\end{cases}
\end{equation}
For $R$ to be a good reaction coordinate, it is required that $P_{unfold}\approx \frac{1}{2}$. The theory has been applied to hairpins and multi-state proteins.  
Using experimentally determinable values for  $s$, one can assess whether  $R$ is a good reaction coordinate  by calculating $P_{unfold}$ using theory.  Applications of the theory to DNA hairpins \cite{Morrison11PRL} showed (Fig. \ref{Tensegrity.fig}C) that the precise sequence determines whether $R$ can be reliably used as an appropriate reaction coordinate, thus establishing the usefulness of the molecular tensegrity parameter.
Application of the theory based on molecular tensegrity showed that $R$ is not a good reaction coordinate for riboswitches \cite{lin2012JPCL}.

\section{Multidimensionality of energy landscape coupled to "memory" affects the force dynamics}
The natural one dimensional reaction coordinate in  mechanical unfolding experiments is the extension $x$ of the molecule.  
In studies of chemistry and biophysics, it is quite standard to project the dynamics of a molecule with many degrees of freedom onto a one-dimensional (1D) reaction coordinate. 
Especially, in single molecule force experiments, the molecular extension ``$x$'', parallel to the direction of an external force, is routinely employed to describe the force-induced dynamics of biomolecules.
However, the assumption of 1D projection is valid only if time scales of dynamics is clearly separated between a variable associated with the reaction coordinate and other degrees of freedom \cite{ZwanzigBook}. 
For a complex biomolecule that has multiple time scales intertwined in its dynamics,  
the signature of interference between two variables with comparable time-scales  
should be displayed in experimental measurements.
As a simple extension of the theories discussed above, we consider a situation of an unbinding process where 
the fluctuation of a molecular surface between open and closed states can gate the unbinding kinetics of a ligand from its binding pocket (Fig.2).  
Depending on the ratio $k/\lambda$ where $\lambda$ is the rate of gating and $k$ is the time scale for unbinding in the absence of gating, the ligand is expected to undergo disparate unbinding kinetics. 
If $k/\lambda\gg 1$ or $k/\lambda\ll1$, the environment appears \emph{static} to the ligand \cite{Zwanzig92JCP,Zwanzig90ACR}. 
Thus, the ligand unbinding occurs via parallel paths over multiple barriers ($k/\lambda\gg 1$) or via single path over a rapidly averaging barrier ($k/\lambda\ll1$).  
Whereas, if $k/\lambda\sim \mathcal{O}(1)$, the open$\leftrightarrow$closed gating produces a fluctuating environment along the dynamic pathway of the ligand and affects the unbinding process in a non-trivial fashion, which is often termed \emph{dynamical disorder}. 
Suppose that the reaction rate constant associated with barrier crossing is controlled by the openness of the receptor molecule, which can be modeled as $kr^2$ where $r$ is the radius of the bottleneck. 
Provided that the distribution of bottleneck size is Gaussian, i.e., $\varphi(r)\sim e^{-r^2/2\theta}$ with $\theta=\langle r^2\rangle$, and $r$ obeys a stochastic differential equation $\partial_tr=-\lambda r+\xi_r(t)$ with $\langle\xi_r(t)\xi_r(t')\rangle=2\lambda\theta\delta(t-t')$, 
two disparate unbinding kinetics are quantified. 
If $\lambda\rightarrow\infty$ much greater than barrier crossing rate then the reactivity is defined by a pre-averaged reaction constant with respect to $r$, i.e., $kr^2\rightarrow k\theta$. 
In this case, the survival probability of the ligand decays exponentially ($S(t)=e^{-k\theta t}$) with $k=k_0e^{fx^{\ddagger}/k_BT}$. 
In contrast, if $\lambda\rightarrow 0$ ($\lambda\ll k$) so that the memory of the initial $r$ value is retained while barrier crossing takes place, $S(t)=\int^{\infty}_0e^{-kr^2t}\varphi(r)dr\sim (1+2k\theta t)^{-1/2}$. 
Two scenarios of bottleneck gating frequency lead to totally different consequences.  
The gating frequency $\lambda$, intrinsic to a molecule of interest, is difficult to adjust independently although it can in principle be varied to a certain extent by changing viscosity.  
However, in single molecule force spectroscopy it is feasible to vary $k$ by controlling the external tension, so that the ratio $k/\lambda$ can be scanned from $k/\lambda\gg 1$ to $k/\lambda\ll 1$ via $k/\lambda\sim \mathcal{O}(1)$. 

The physical picture described above can be mathematically formulated as follows. 
In the context of ligand binding to myoglobin Zwanzig first proposed such a model by assuming 
that  reaction (binding) takes place along the $x$-coordinate and at the barrier top ($x=x_{ts}$) the reactivity is determined by the cross section of bottleneck described by the $r$-coordinate \cite{Zwanzig92JCP}. 
We adopt a similar picture to describe the modifications when such a process is driven by force. 
First, consider the Zwanzig case i.e, with loading rate $r_f = 0$.
The equations of motion for $x$ and $r$ are, respectively, 
\begin{eqnarray}
m\frac{d^2x}{dt^2}&=&-\zeta\frac{dx}{dt}-\frac{dU(x)}{dx}+F_x(t)\nonumber\\
\frac{dr}{dt}&=&-\gamma r+F_r(t).
\label{eqn:eqofmotion}
\end{eqnarray}
The Liouville theorem ($\frac{d\rho}{dt}=0$) describes the time evolution of probability density, $\rho(x,r,t)$, as 
\begin{equation}
\frac{d\rho}{dt}=\frac{\partial\rho}{\partial t}+\frac{\partial}{\partial x}\left(\frac{dx}{dt}\rho\right)+\frac{\partial}{\partial r}\left(\frac{dr}{dt}\rho\right)=0.
\label{eqn:Liouville}
\end{equation}
 By inserting of Eq.\ref{eqn:eqofmotion} to Eq.\ref{eqn:Liouville} and  neglecting the inertial term, ($m\frac{d^2x}{dt^2}$), 
and averaging over the white-noise spectrum, 
and the fluctuation-dissipation theorem
($\langle F_x(t)F_x(t')\rangle=2\zeta k_BT\delta(t-t')$, $\langle F_r(t)F_r(t')\rangle=2\lambda\theta\delta(t-t')$ 
where $\langle r^2\rangle\equiv \theta$) leads to a Smoluchowski equation 
for $\rho(x,r,t)$ in the presence of a reaction sink. 
\begin{equation}
\frac{\partial\overline{\rho}}{\partial t}=\mathcal{L}_x\overline{\rho}+\mathcal{L}_r\overline{\rho}-k_rr^2\delta(x-x_{ts})\overline{\rho}
\label{eqn:Smol}
\end{equation} 
where $\mathcal{L}_x\equiv D\frac{\partial}{\partial x}\left(\frac{\partial}{\partial x}+\frac{1}{k_BT}\frac{dU(x)}{dx}\right)$ and $\mathcal{L}_r\equiv \lambda\theta\frac{\partial}{\partial r}\left(\frac{\partial}{\partial r}+\frac{r}{\theta}\right)$. 
Integrating  both sides of Eq.\ref{eqn:Smol} using $\int dx\rho(x,r,t)\equiv \overline{C}(r,t)$ leads to 
\begin{equation}
\frac{\partial\overline{C}}{\partial t}=\mathcal{L}_r\overline{C}(r,t)-k_rr^2\overline{\rho}(x_{ts},r,t). 
\label{eqn:step1}
\end{equation}
By writing 
$\overline{\rho}(x_{ts},r,t)=\phi_x(x_{ts})\overline{C}(r,t)$ 
where $\phi(x_{ts})$ should be constant as 
$\phi(x_{ts})=e^{-U(x_{ts})/k_BT}/\int dx e^{-U(x)/k_BT}\approx \sqrt{\frac{U''(x_b)}{2\pi k_BT}}e^{-(U(x_{ts})-U(x_b))/k_BT}$, 
 Eq.\ref{eqn:step1} becomes
\begin{equation}
\frac{\partial\overline{C}}{\partial t}=\mathcal{L}_r\overline{C}(r,t)-kr^2\overline{C}(r,t). 
\label{eqn:step2}
\end{equation}
where $k\equiv k_r\sqrt{\frac{U''(x_b)}{2\pi k_BT}}e^{-(U(x_{ts})-U(x_b))/k_BT}$. In all likelihood $k_r$ reflects the dynamics near the barrier, 
so we can write $k=\kappa \frac{\omega_{ts}\omega_b}{2\pi\gamma}e^{-\Delta U/k_BT}$ where $\kappa$ describes the geometrical information of the cross section of bottleneck. Now we rederive the equation in Zwanzig's seminal paper where the 
survival probability ($\Sigma(t)=\int^{\infty}_0 dr\overline{C}(r,t)$) is given under a reflecting boundary condition at $r=0$ and Gaussian initial condition $\overline{C}(r,t=0)\sim e^{-r^2/2\theta}$. 
By setting $\overline{C}(r,t)=\exp{(\nu(t)-\mu(t)r^2)}$, Eq.\ref{eqn:step2} can be solved exactly, leading to 
\begin{eqnarray}
\nu'(t)&=&-2\lambda\theta\mu(t)+\lambda\nonumber\\
\mu'(t)&=&-4\lambda\theta\mu^2(t)+2\lambda\mu(t)+k.
\end{eqnarray}
The solution for $\mu(t)$ is obtained by solving 
$\frac{4\theta}{\lambda}\int^{\mu(t)-1/4\theta}_{1/4\theta}\frac{d\alpha}{\sigma^2-16\theta^2\alpha^2}=t$, and this leads to 
\begin{eqnarray}
\frac{\mu(t)}{\mu(0)}&=&\frac{1}{2}\left\{1+S\frac{(S+1)-(S-1)E}{(S+1)+(S-1)E}\right\}\nonumber\\
\nu(t)&=&-\frac{\lambda t}{2}(S-1)+\log{\left(\frac{(S+1)+(S-1)E}{2S}\right)^{-1/2}}
\end{eqnarray}
with $\mu(0)=1/2\theta$.
The survival probability, which was originally obtained by Zwanzig,  is 
\begin{equation}
\Sigma (t)=\exp{\left(-\frac{\lambda}{2}(S-1)t\right)}
\left(\frac{(S+1)^2-(S-1)^2E}{4S}\right)^{-1/2}
\label{eqn:solution}
\end{equation}
where $S=\left(1+\frac{4k\theta}{\lambda}\right)^{1/2}$ and $E=e^{-2\lambda St}$.

We wish to examine the consequences of coupling between local and global reaction coordinates under tension. In order to accomplish our goal we solve the Smoluchoski equation in the presence of constant loading rate. In this case, 
$k$ in Eq.\ref{eqn:step2} should be replaced with $ke^{t (r_f\Delta x^{\ddagger}/k_BT)}$. 
Eq.\ref{eqn:step2}, however, becomes hard to solve if the sink term depends on $t$. 
Nevertheless, analytical solutions can be obtained for special cases of $\lambda$. 
If $\lambda\rightarrow \infty$, $d\Sigma(t)/dt=-k\theta e^{t(r_f\Delta x^{\ddagger}/k_BT)}\Sigma(t)$, and hence, 
\begin{equation}
\Sigma(f)=\exp{\left(-k\theta\int^f_0df\frac{1}{r_f}e^{f\Delta x^{\ddagger}/k_BT}\right)}=\exp{\left(-\frac{k\theta k_BT}{r_f\Delta x^{\ddagger}} (e^{f\Delta x^{\ddagger}/k_BT}-1)\right)}
\end{equation}
Using the  rupture force distribution $P(f)=-\frac{d\Sigma(f)}{df}$ and $\frac{dP(f)}{df}|_{f=f^*}=0$, one can obtain 
the most probable force 
\begin{equation}
f^*=\frac{k_BT}{\Delta x^{\ddagger}}\log{\frac{r_f\Delta x^{\ddagger}}{(k\theta)k_BT}}.
\end{equation}

If $\lambda$ is small ($\lambda\rightarrow 0$) then 
$\overline{C}(r,t)=\exp{[-\int_0^t dt kr^2\exp{(t\times r_f\Delta/k_BT)}]}=\exp{\left[-\frac{kr^2 k_BT}{r_f\Delta x^{\ddagger}}(e^{tr_f\Delta x^{\ddagger}/k_BT}-1)\right]}$ with the initial distribution of $e^{-r^2/2\theta}$.
Thus,
\begin{eqnarray}
\Sigma(f)&=& \int_0^{\infty}dr\exp{\left[-r^2\frac{kk_BT}{r_f\Delta x^{\ddagger}}(e^{f\Delta x^{\ddagger}/k_BT}-1)\right]}\sqrt{\frac{2}{\pi\theta}}\exp{\left[-r^2/2\theta\right]}\nonumber\\
&=& \sqrt{\frac{2}{\pi\theta}}\left(\frac{kk_BT}{r_f\Delta x^{\ddagger}}(e^{f\Delta x^{\ddagger}/k_BT}-1)+\frac{1}{2\theta}\right)^{-1/2}. 
\end{eqnarray}
Note that if $r_f\rightarrow 0$ we recover Zwanzig's result $\Sigma(t)\sim (1+2k\theta t)^{-1/2}$. Using $P(f)=-d\Sigma(f)/df$
\begin{equation}
P(f)=\frac{1}{\sqrt{2\pi\theta}}\left(\frac{kk_BT}{r_f\Delta x^{\ddagger}}(e^{f\Delta x^{\ddagger}/k_BT}-1)+\frac{1}{2\theta}\right)^{-3/2}\frac{k}{r_f}e^{f\Delta x^{\ddagger}/k_BT}. 
\end{equation}
$dP(f)/df|_{f=f^*}=0$ gives 
\begin{equation}
f^*=\frac{k_BT}{\Delta x^{\ddagger}}\log{\left\{\left(\frac{r_f\Delta x^{\ddagger}}{k\theta k_BT}\right)\left(1-\frac{2k\theta k_BT}{r_f\Delta x^{\ddagger}}\right)\right\}}, 
\label{eqn:smalllambda}
\end{equation}
in which $r_f\geq(1+2\theta)\frac{k\theta k_BT}{\Delta x^{\ddagger}}$ since $f^*\geq 0$.
This shows that $f^*$ vs $r_f$ has a different form when $\lambda\rightarrow 0$ from the one when $\lambda\rightarrow\infty$.
The deviation of Eq.\ref{eqn:smalllambda} from the conventional relation is pronounced when $\left[r_f-(1+2\theta)\frac{kk_BT}{\Delta x^{\ddagger}}\right]\rightarrow 0^+$.

At present, experimental data have been interpreted using mostly one dimensional free energy profiles.  
The meaning and the validity of the extracted free energy profiles has not been established. 
At the least, this would require computing the force-dependent first passage times using the "experimental" 
free energy profile assuming that the extension is the only slowly relaxing variable. 
If the computed force-dependent rates (inversely proportional to first passage times) agree with the measured rates then the use of extension of the reaction coordinate would be justified.  
In the absence of good agreement with experiments other models, such as the one we have proposed here, must be considered.  
In the context of force-quench refolding we have shown (see below) that extension alone is not an adequate reaction coordinate \cite{Hyeon08JACS}.   
For refolding upon force-quench of RNA hairpins, the coupling between extension and local dihedral angles, which reports on the conformation of the RNA, needs to be taken into account to quantitatively describe the refolding rates.    

\section{Conclusions}
With the advent of single molecule experiments that can manipulate biomolecules using mechanical force it has become possible to characterize energy landscapes quantitatively. Mechanical folding and unfolding trajectories of proteins and RNA show that there is great diversity in the explored routes \cite{Hyeon06Structure,Bustamante03Science,Fernandez04Science}.   In certain well defined systems with simple native states, such as RNA and DNA hairpins, it has been shown using constant force unfolding that the hairpins undergo sharp bistable transitions from folded to unfolded states \cite{Bustamante02Science,Block06Science,Woodside06PNAS}.  
From the dynamics of the extension as a function of time measured over a long period the underlying force dependent profiles have been inferred.  
The force-dependent folding and unfolding rates and the unfolding trajectories can be used to construct the one-dimensional energy landscape. In a remarkable paper \cite{Block06Science}, Block and coworkers have shown that the location of the TS can be moved, at will, by varying the hairpin sequence.   
The TS was obtained using the Bell model by assuming that the $\Delta x^{\ddagger}$ is independent of $f$.  
While this seems reasonable given the sharpness of the inferred free energy profiles near the barrier top it will be necessary to show the $\Delta x^{\ddagger}$ does not depend on force.

The fundamental assumption in inverting the force-clamp data is that the molecular extension is a suitable reaction coordinate.  This may indeed be the case for force-spectroscopy in which the  response of the molecule only depends on force that is coupled to the molecular extension, which may well  
represent the slow degrees of freedom.  
The approximation is more reasonable for forced-unbinding. 
It is less clear if it can be assumed that extension $x$ is the appropriate reaction coordinate when refolding is initiated by quenching the force to low enough values such that the folded state is preferentially populated.  
In this case the dynamic reduction in $x$  can be coupled to collective internal degrees of freedom. 
In a recent paper \cite{Hyeon06BJ} we showed, in the context of force-quench refolding of an RNA hairpin, that the reduction in $x$ is largely determined by local conformational changes in the dihedral angle degrees in the loop region.  Zipping by nucleation of the hairpin with concomitant reduction in $x$ does not occur until the transitions from \textit{trans} to \textit{gauche} state in a few of the loop dihedral angles take  place.  
In this case, one has to consider at least a two dimensional free energy landscape. 
Fig.\ref{multidimension_RNA.fig} clearly shows such a coupling between end-to-end distance ($R$) and the dihedral angle degrees of freedom. The "correctness" of the six dihedral angles representing the conformation of the RNA hairpin loop region ($\phi_i$, $i=19,\ldots 24$) is quantified using $\langle 1-\cos{(\phi_i-\phi_i^o)}\rangle$, where $\phi^o_i$ is the angle value in the native state and $\langle\ldots\rangle$ is the average over the six dihedral angles. 
$\langle 1-\cos{(\phi-\phi^o)}\rangle=0$ signifies the correct dihedral conformation for the hairpin loop region. 
Once the "correct" conformation is attained in the loop region, the rest of the zipping process can easily proceed as we have shown in \cite{Hyeon06BJ}. 
Before the correct loop conformation is attained, RNA spends substantial time in searching the conformational space related to the dihedral angle degree of freedom. 
The energy landscape ruggedness is manifested as in Fig.\ref{multidimension_RNA.fig} when conformational space is represented using multidimensional order parameters.   The proposed coupling between the local dihedral angle degrees of freedom and extension (global parameter) is fairly general.  A similar structural slowing down, due to the cooperative link between local and global coordinates, should be observed in force-quench refolding of proteins as well.

One of the most exciting uses of singe molecule experiments is their ability to extract precise values of the energy landscape roughness $\epsilon$ by using temperature as a variable in addition to $f$.  
In this case a straightforward measurement of the unbinding rates as a function of $f$ or $r_f$ can be used to obtain $\epsilon$ without having to make \textit{any assumptions} about the underlying mechanisms of unbinding.  
Of course, this involves performing a number of experiments. 
In so doing one can also be rewarded with a diagram of states in terms of $f$ and $T$ \cite{Hyeon05PNAS}.  
The theoretical calculations and arguments given here also show that the power of single molecule experiments can be fully realized only by using data from the experiments in conjunction with  carefully designed theoretical and computational models \cite{Hinczewski13PNAS}.  
The latter can provide the structures that are  sampled in the process of forced-unfolding and force-quench refolding as was illustrated for ribozymes and GFP \cite{Hyeon06Structure}.  
It is likely that the promise of measuring the energy landscapes of biomolecules, one molecule at a time, will be fully realized using a combination of single molecule measurements, theory, and simulations.  Recent studies have already given us a glimpse of that promise with more to come shortly.


\clearpage

\section*{\bf Figure Caption}

\begin{figure}[ht]
\includegraphics[width=3.0in]{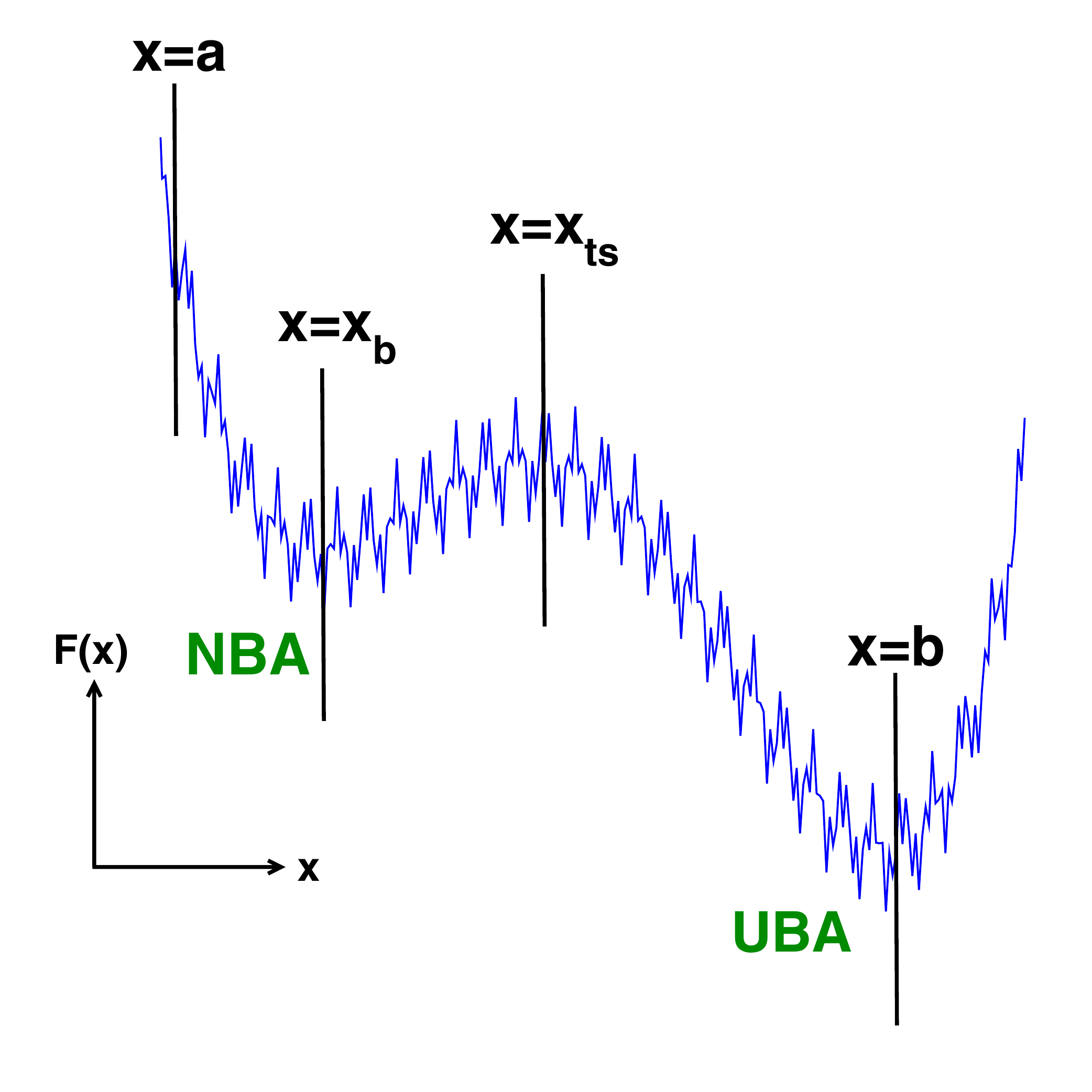}
\caption{\label{landscape}}
\end{figure}

{\bf Figure \ref{landscape} :}
Caricature of the rough energy landscape of proteins and RNA that fold in an apparent ''two-state'' manner using extension $x$, the coordinate that is
conjugate to force $f$. 
Under force $f$, the zero-force free energy profile ($F_0(x)$) is tilted 
by  $f\times x$ and gives rise to the free energy profile, $F(x)$. 
In order to clarify the derivation of Eq.\ref{eqn:mfpt} we have explicitly indicated the average location of the relevant parameters.\\

\begin{figure}[ht]
\includegraphics[width=6.0in]{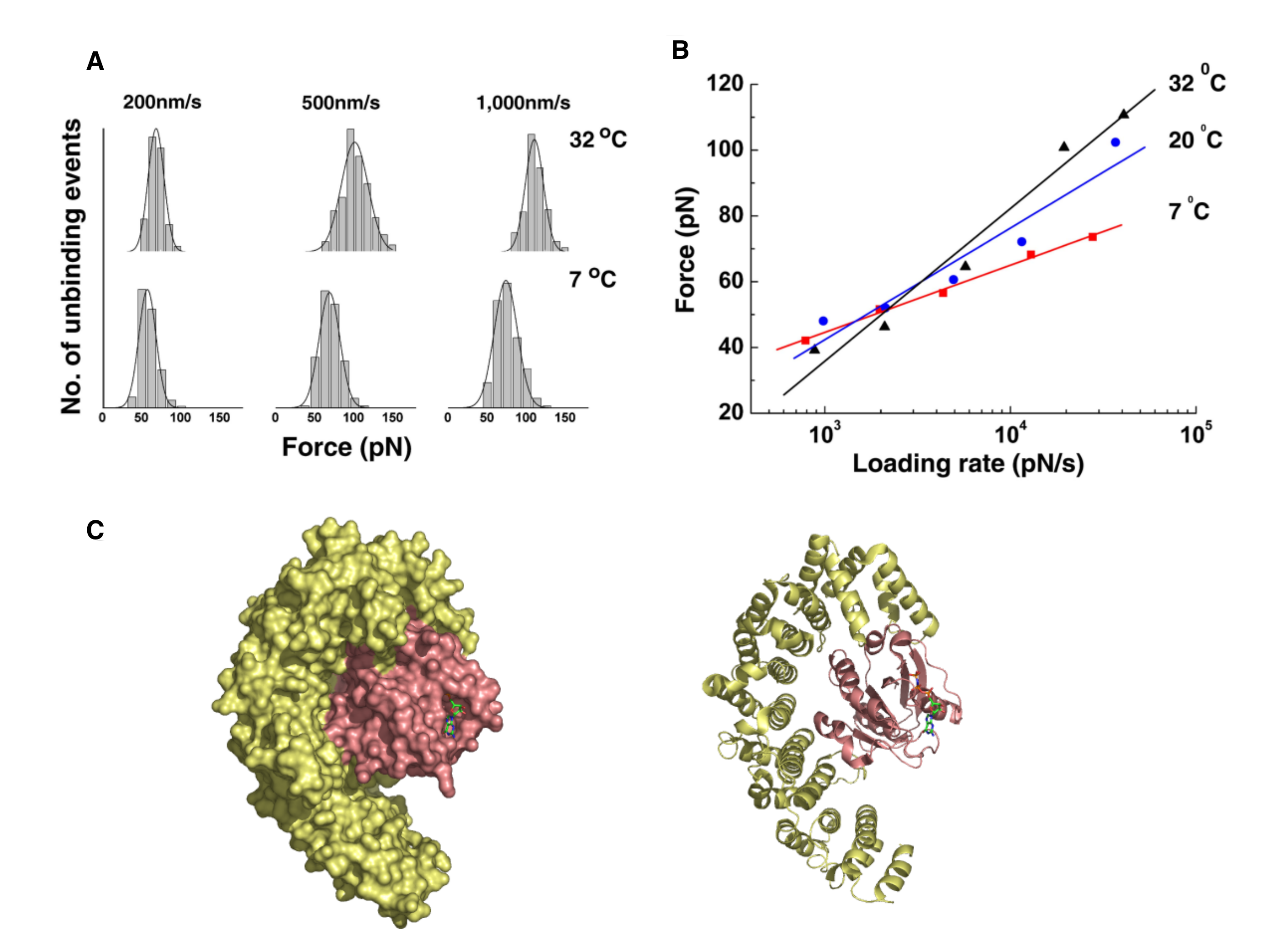}
\caption{\label{Nevo_fig}}
\end{figure}

{\bf Figure \ref{Nevo_fig} :}
Dynamic force spectroscopy measurements of single imp-$\beta$-RanGppNHp pairs at different temperatures. 
{\bf A.} Distributions of measured unbinding forces using AFM for the lower-strength 
conformation of the complex at different loading rates at 7 and 32 $^o$C. 
Roughness acts to increase the separation between the distributions recorded at different temperatures. The histograms are fit using Gaussian distributions. 
The width of the bins represents the thermal noise of the cantilever. 
{\bf B.} Force spectra used in the analysis. The most probable unbinding forces $f^*$ are plotted as a function of $\log(r_f)$. 
The maximal error is $\pm10$\% because of uncertainities in 
determining the spring constant of the cantilevers. 
Statistical significance of the differences between the slopes of the spectra was confirmed using covariance test. 
(Images courtesy of Reinat Nevo and Ziv Reich \cite{Reich05EMBOrep}). {\bf C.} Ran-importin$\beta$ complex crystal structures (PDB id: 1IBR \cite{Vetter99Cell}) in surface (left) and ribbon (right) representations. In AFM experiments, 
Ran (red) protein complexed to importin$\beta$ (yellow) is pulled until the dissociation of the complex takes place. \\

\begin{figure}[ht]
\includegraphics[width=7.0in]{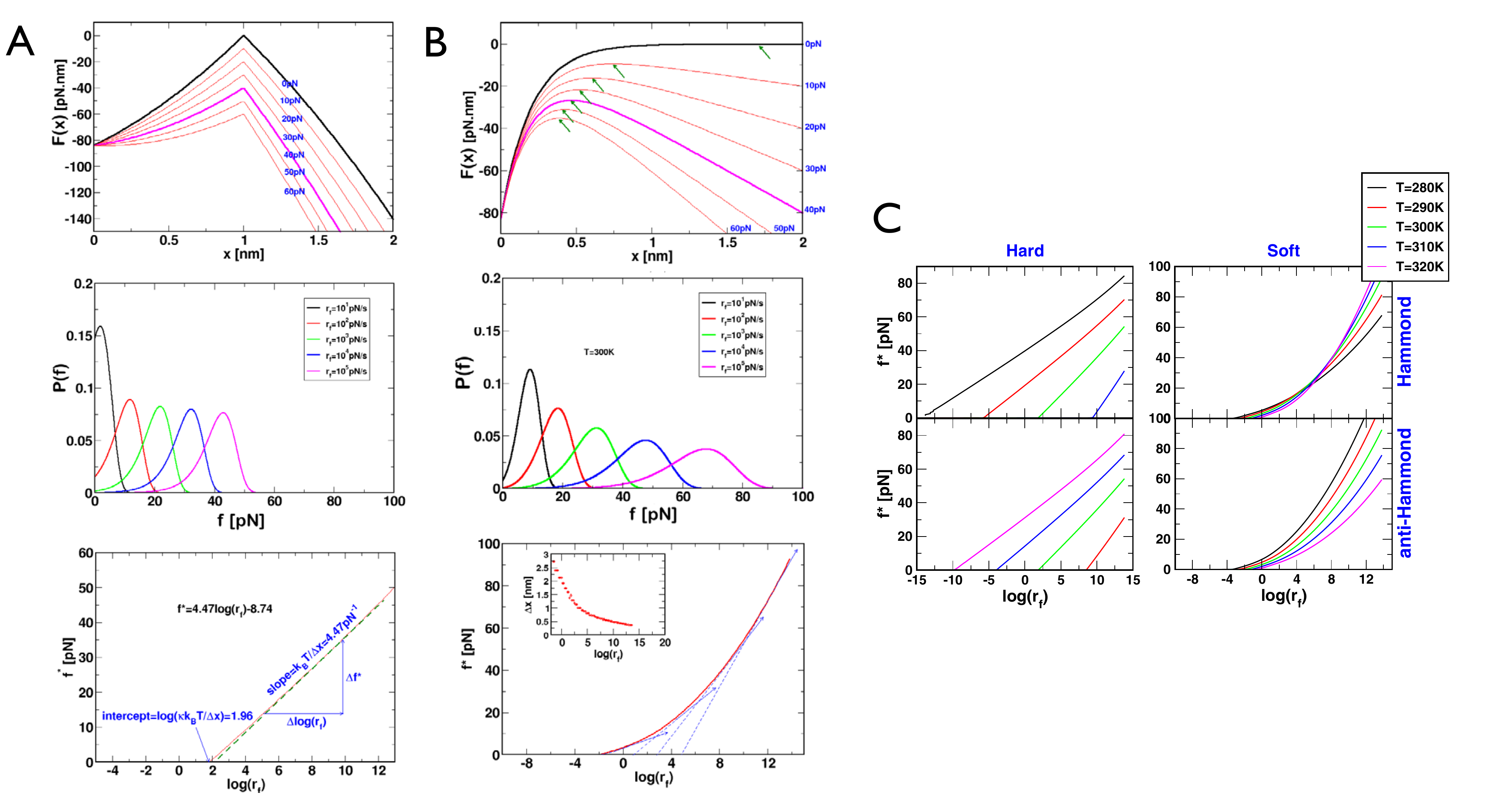}
\caption{\label{hard_soft.fig}}
\end{figure}

{\bf Figure \ref{hard_soft.fig} :}
Dynamic force spectroscopy analysis using two different free energy models.
From the top to bottom, (top) free energy profiles at different forces, (middle) rupture force distributions, and (bottom) relationship of most probable force as a function of log-loading rates are shown for ({\bf A}) hard and ({\bf B}) soft potentials.
For hard potential, free energy profile in the absence of force is $F_0(x)=-V_0|(x+1)^2-\xi^2|$ with $V_0=20$ pN/nm, $\xi=4$ nm, and $x\geq 0$. The lack of change in $x_{ts}$ as $f$ changes shows a \emph{brittle response} under tension. 
For soft potential, free energy profile in the absence of force is $F_0(x)=-V_0\exp{(-\xi x)}$ with $V_0=82.8$ pN$\cdot$nm, $\xi=4$ (nm)$^{-1}$.
The lack of change in $x_{ts}$ as $f$ changes shows a \emph{brittle response} under tension. 
For emphasis on the soft response of the potential, 
the position of TS at each force value is indicated with arrows. 
{\bf C}. By tuning the value of $\xi$ (see Eq. (23)) as a function of temperature, 
Hammond and anti-Hammond behaviors emerge  
in the context of force spectra in the free energy profiles that show hard and soft responses. The condition for Hammond or anti-Hammond
behavior depends on $\alpha$ (Eq. (23)). \\

{\bf Figure \ref{FB.fig}:}
Fluctuating bottleneck model under an external force. Gating dynamics of receptor protein conformation with frequency $\lambda$ can interfere with the dynamics of ligand unbinding if the unbinding rate ($k$) is comparable or slower than $\lambda$.\\

\begin{figure}[ht]
\includegraphics[width=6.0in]{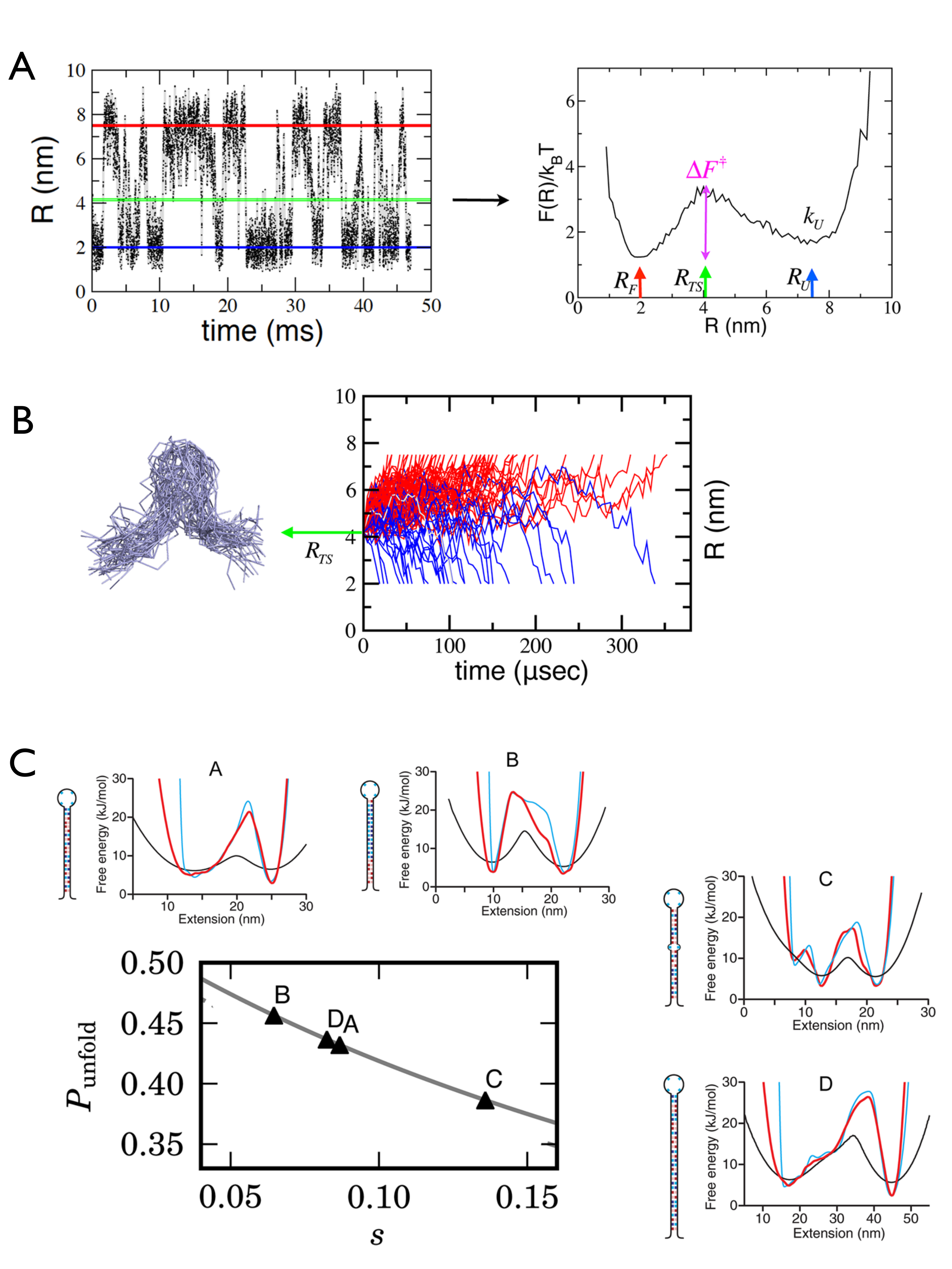}
\caption{\label{Tensegrity.fig}}
\end{figure}

\begin{figure}
 \includegraphics[width=5.0in]{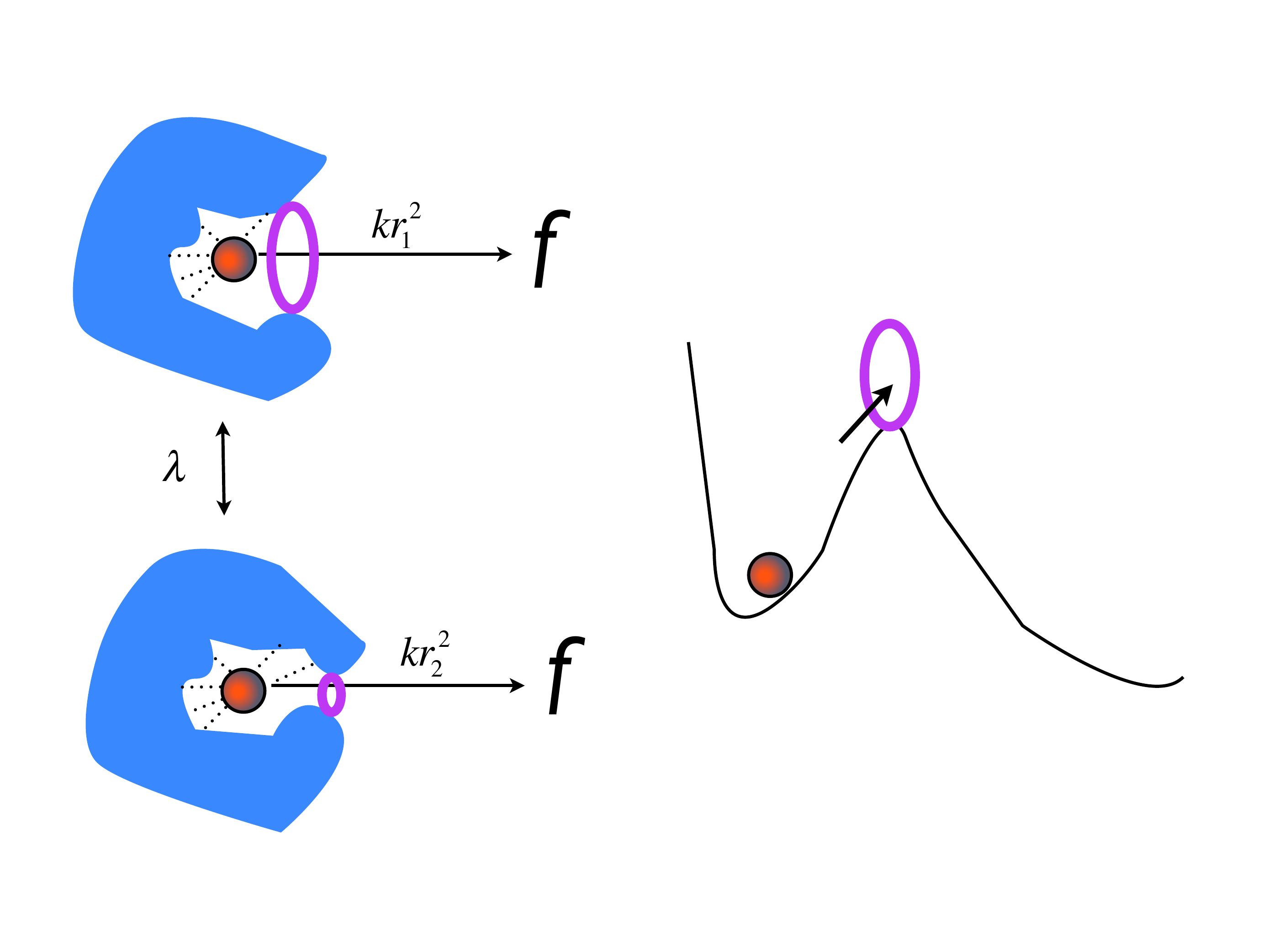}
 \caption{ \label{FB.fig}}
\end{figure}

{\bf Figure \ref{Tensegrity.fig}:}
{\bf A}. Time trace of end-to-end distance ($R$) of RNA hairpin at transition midforce $f_m = 14.7$ pN (left). The corresponding free energy profile in terms of $R$, $F(R)$ (right). 
The positions of native, unfolded, and transition states are marked with arrows. In addition, barrier height ($\Delta F^{\ddagger}$) and the curvature of the unfolded state ($k_U$) are also shown on the $F(R)$. {\bf B}. Time trajectories of simulations starting from the configurations of the transition state ensemble (shown on the left). Trajectories reaching the folded and unfolded state at 2.0 and 7.5 nm are colored in blue and red, respectively. In the blue trajectories, a number of recrossing events can be observed.
{\bf C}. Tensegrity parameters calculated for four DNA hairpins with different sequences in Ref. \cite{Greenleaf08Science} are related to $P_{unfold}$. The DNA hairpin with sequence B is predicted to have $P_{unfold}$ most proximal to 0.5, which suggests that the free energy profile calculated in terms of end-to-end distance coordinate most accurately describes the dynamics of this DNA hairpin.\\

\begin{figure}
 \includegraphics[width=6.5in]{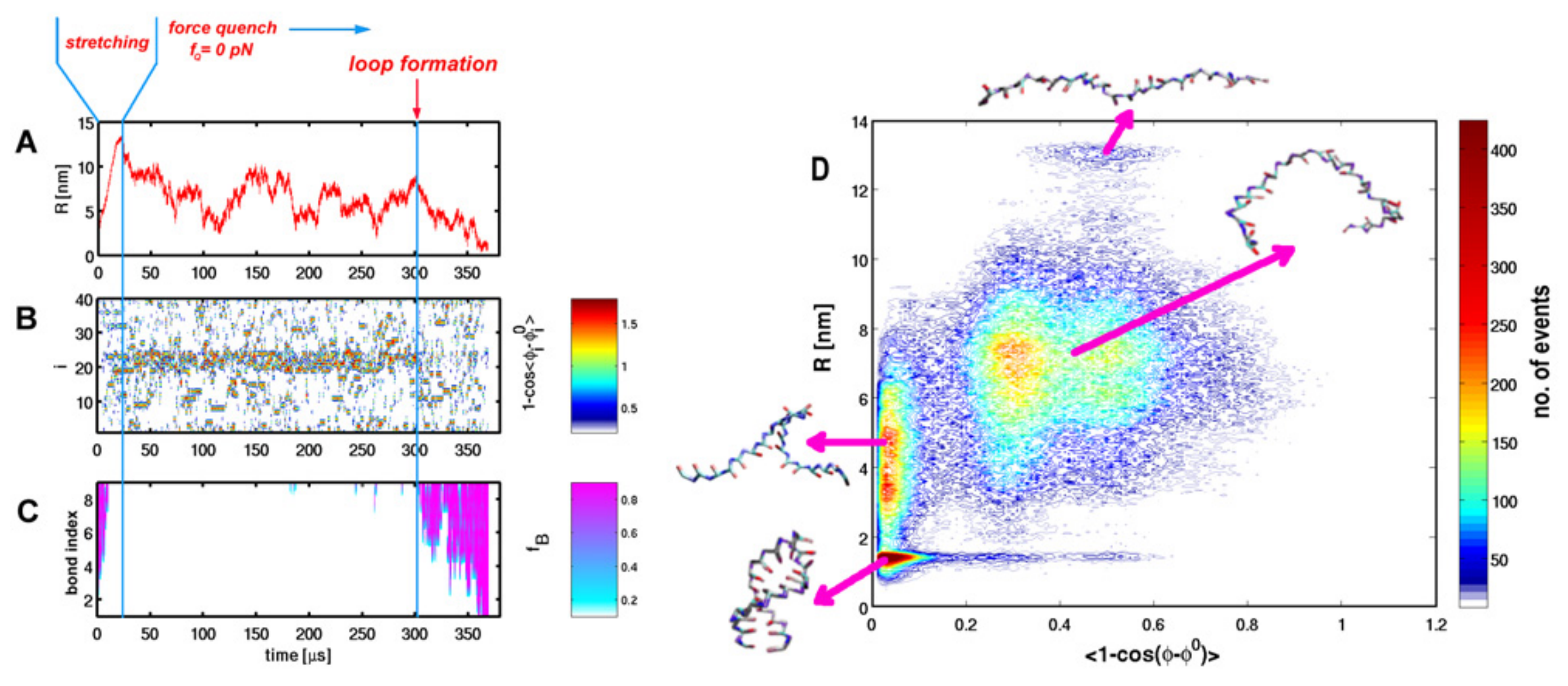}
 \caption{\label{multidimension_RNA.fig}}
\end{figure}

{\bf Figure \ref{multidimension_RNA.fig} :}
{\bf A.} A sample refolding trajectory of a RNA hairpin  starting from the stretched state.
The hairpin was, at first, mechanically unfolded to a fully stretched state and the force was subsequently quenched
 to zero at $t\approx 20$ $\mu s$. The time-dependence of the
end-to-end distance shows that force-quench refolding occurs in steps.
{\bf B.} The deviation of the dihedral angles from their values in the native state as a function of  time shows  large departures from native values  of
the dihedral angles in loop region (indicated by the red strip).
Note that this strip disappears around $t\approx 300$ $\mu s$, which coincides with the formation of bonds
shown in {\bf C}. $f_B$ is the fraction of bonds in pink that  indicates that the bond is fully formed.
{\bf D.} The histograms collected from the projections of twelve stretching and force-quench refolding trajectories on the two dimensional plane characterized by the end-to-end distance ($R$) and the average correctness of dihedral angles ($\langle 1-\cos{(\phi-\phi^0)}\rangle$) around the loop region ($i=19-24$).  The scale on the right gives the density of points in the two dimensional projection.  This panel shows that the local dihedral angles are coupled to the end-to-end distance $R$, and hence extension alone is not a good reaction coordinate especially in force-quench refolding. The molecular extension $x$ is related to $R$ by $x = R -R_N$ where $R_N$ is the distance in the folded state.

\clearpage

\end{document}